\documentclass{elsart}

\usepackage{graphicx}
\usepackage{amsmath}

\begin{document}
\begin{frontmatter}
\title{Nonstationary excitations in Bose-Einstein condensates
under the action of periodically varying scattering length with
time dependent frequencies}

\author{S.~Rajendran},
\author{P.~Muruganandam}, and
\author{M.~Lakshmanan\corauthref{cor1}}
\ead{lakshman@cnld.bdu.ac.in}
\corauth[cor1]{Corresponding Author.  Fax:+91-431-2407093}
\address{Centre for Nonlinear Dynamics, Department of Physics,
Bharathidasan University, Tiruchirappalli - 620 024, India}
\date{}

\begin{abstract}

We investigate nonstationary excitations in 3D-Bose-Einstein
condensates in a spherically symmetric trap potential under the
modulation of scattering length with slowly varying frequencies
(adiabatic modulation). By numerically solving the Gross-Pitaevskii
equation we observe a step-wise increase in the amplitude of
oscillation due to successive phase locking between driving
frequency and nonlinear frequency. Such a nonstationary excitation
has been shown to exist by an analytic approach using variational
procedure and perturbation theory in the action-angle variables. By
using a canonical perturbation theory, we have identified  the
successive resonance excitations whenever the driven frequency
matches the nonlinear frequency or its subharmonics.
\end{abstract}

\end{frontmatter}
\section{Introduction}

The successful experimental realization of trapped Bose-Einstein
condensates in alkali metal atoms has triggered immense interest in
understanding the various properties of the ultra cold
matter~\cite{Ketterle:95-01,Wieman_cornell:96-01}. The properties of
the condensate wave function is usually described by a mean field
Gross-Pitaevskii equation~\cite{Dalfovo_etal}. For the past couple
of years, there has been increased interest in studying the
properties of Bose-Einstein condensates with time varying trap
potentials and scattering lengths both experimentally and
theoretically~\cite{Inouye_etal,Staliunas_etal:2002,Saito_ueda:prl2003,Abdul_etal:pra2003:1,Ska_pra:04-01}.
In particular, temporal periodic modulation of scattering length by
exploiting a Feshbach resonance is given a central importance in
recent times. Earlier studies show that, in certain circumstances,
periodically varying scattering length can stabilize the collapsing
condensate~\cite{Towers_Malomed:josab:02:01,Saito_ueda:prl2003,Abdul_etal:pra2003:1,Ska_pra:04-01}.
However, very recently it has been shown that for a sign alternating
nonlinearity an increase in the frequency of oscillations
accelerates collapse~\cite{Konotop_pacciani:05:01}.

From another point of view, these periodic modulations lead to
resonance phenomena in the condensates. Resonance is an
interesting feature of an oscillation under the action of an
external periodic force manifesting in a large amplitude, when the
frequency of the external force equals an integral multiple of the
natural frequency of oscillation. Although, the phenomenon of
resonance is well understood in linear systems, the nonlinearity
that arises in the Bose-Einstein condensates due to the
inter-atomic interactions leads to an important problem of
nonlinear resonance of wide interest.

In general, there are two ways of driving a nonlinear oscillator by
small driving periodic force, namely, external and parametric. If
the driving frequency is constant (that is, driving force is exactly
periodic with time), the initial growth of the oscillator's
amplitude with time is arrested by its nonlinearity. On the other
hand, if the driving frequency is slowly varying with time (driving
force is almost periodic with time), the oscillator can stay phase
locked but, on an average, increase its amplitude with time or a
persistent growth in the amplitude takes place, and this phenomenon
is known as \emph{autoresonance}. In such systems, the nonlinear
frequency is slowly varying with time. This autoresonance phenomena
is also referred as adiabatic nonlinear phase locking and
synchronization. However, in certain nonlinear systems, frequency is
independent of the amplitude of oscillation
\cite{Vkc:2005,Vkc_proc:2006}. In these systems, when the driving
frequency is slowly varying with time, one might expect successive
resonance excitations at subharmonic frequencies. We refer this type
to resonance as a kind of subharmonic autoresonance
\cite{Friedland:00-01}.

In the present work, we identify such a subharmonic autoresonance in
Bose-Einstein condensates where it shows successive resonance
excitations due to periodic modulation with slowly varying
frequency. In Bose-Einstein condensates it is experimentally
feasible to vary the scattering length by either magnetically or
optically inducing a Feshbach resonance
\cite{Inouye_etal,Fatemi_jones_lett:2000}. Earlier works on
Bose-Einstein condensation report the periodic modulation with
constant frequency of scattering length
\cite{Staliunas_etal:2002,Saito_ueda:prl2003,Abdul_etal:pra2003:1,Ska_pra:04-01}.
Along these lines, it is of potential interest to understand the
dynamics of Bose-Einstein condensates under the action of
periodically varying scattering length with slowly varying
frequency. Motivated by the above, in this paper, we study the
effect of such an excitation on the 3D Bose-Einstein condensates in
a spherically symmetric trap potential and investigate the nature of
parametric resonance. In particular, we point out the step-wise
increase in amplitude of oscillation due to successive phase locking
between the driving frequency and the nonlinear frequency (or its
subharmonics) of the system.

This paper is organized as follows. In Sec.~\ref{gpe} we discuss
the properties of Bose-Einstein condensates driven by a periodic
force with time varying frequency from numerical simulations
using pseudo-spectral method. In Sec.~\ref{variational} we
describe the variational procedure and derive a reduced system of
ordinary differential equations (ODEs) to describe the dynamics of
condensate width. In sec.~\ref{action} we analyze the width
dynamics using a perturbed action angle variable theory for the
reduced system of ODEs. Then in Sec.~\ref{canonical}, by applying
a canonical perturbation theory, we deduce the approximate
nonlinear frequencies which are responsible for successive
resonance excitation in BEC under the periodic modulation with
slowly varying frequency. Finally in Sec.~\ref{conclusion}, we
give a summary and conclusions.

\section{Nonlinear Gross-Pitaevskii Equation}
\label{gpe}

At ultra low temperatures, the time-dependent Bose-Einstein
condensate wave function $\Psi({\bf r}; \tau)$ at position ${\bf
r}$ and time $\tau $ may be described by the following  mean-field
nonlinear GP equation~\cite{Dalfovo_etal},
\begin{align}
\left[ -\frac{\hbar^2\nabla ^2}{2m} + V({\bf r}) + \hat g
N\vert\Psi({\bf r};\tau)\vert^2 -i\hbar\frac{\partial }{\partial
\tau} \right]\Psi({\bf r};\tau)=0. \label{a}
\end{align}
Here $m$ is the mass and  $N$ is the number of atoms in the
condensate, $\hat g = 4\pi \hbar^2 \tilde a/m $ is the strength of
interatomic interaction, with $\tilde a$ being the periodically
varying atomic scattering length. The normalization condition of the
wave function is $\int d{\bf r}\vert \Psi({\bf r};\tau)\vert^2 = 1$.
The three-dimensional trap potential is given by $V({\bf r}) =
\frac{1}{2}m \left(\omega_x^2\bar x^2 + \omega_y^2\bar y^2+\omega_
z^2\bar z^2\right)$, where $\omega_x$, $\omega_y$, and $\omega_z$
are the angular frequencies in the $\bar x$, $\bar y$ and $\bar z$
directions, respectively, and ${\bf r}\equiv (\bar x,\bar y,\bar z)$
is the radial vector (The standard notation ($x$,$y$,$z$) is
reserved below for rescaled variables).

\subsection{Spherically symmetric trap potential}

In the spherically symmetric trap, i.e., $\omega_x = \omega_y =
\omega_z \equiv \omega_0$ the trap potential is given by $V({\bf
r})=\frac{1}{2}m\omega_0 ^2\tilde{r}^2$, where $\omega_0$ is the
angular frequency and $\tilde{r}$ the radial distance. The wave
function can be written as $\Psi({\bf r};\tau)=\psi(\tilde
r,\tau)$. After a transformation of variables to dimensionless
quantities defined by $r =\sqrt 2\tilde r/l$, $t=\tau \omega_0$,
$l\equiv \sqrt {(\hbar/m\omega_0)} $ and $\phi(r,t) \equiv
\varphi(r,t)/r  = \psi(\tilde r,\tau)(4\pi l^3)^{1/2}$, the GP
equation for the ground state wave function becomes
\begin{align}
\left[-\frac{\partial^2} {\partial r^2}+\frac{r^2}{4}+g\left|
\frac{\varphi(r,t)}{r} \right| ^2 -i\frac{\partial }{\partial
t}\right] \varphi (r,t)=0, \label{gpe_r}
\end{align}
where $g = N \tilde a/l$. The normalization condition for the wave
function is then
\begin{align}\label{n1}
\int_0^{\infty} \vert \varphi(r,t) \vert^2\,dr = 2 \sqrt{2}.
\end{align}

\subsection{Numerical results}

In order to study the nature of the condensate wave function, we
solve the GP equation~(\ref{gpe_r}) numerically using a
pseudo-spectral method~\cite{Pm_ska_iitkg:2003}. In the
pseudo-spectral method the condensate wave function is expanded in
terms interpolating polynomials. When this expansion is substituted
into the GP equation, the (space) differential operators operate on
a set of known polynomials and generate a differentiation matrix
operating on the unknown coefficients. Consequently, the
time-dependent partial-differential nonlinear GP equation in space
and time variables is reduced to a set of coupled ordinary
differential equations in time which can then be solved by a
fourth-order adaptive step-size controlled Runge-Kutta
method~\cite{Press_etal}.

A similar pseudo-spectral method has been used in ~\cite{Ska_pm_jpb:2}
with Hermite polynomials as the interpolant for the case of
completely anisotropic trap potential. However, for the
spherically symmetric condensates described by Eq.~(\ref{gpe_r}),
it will be more advantageous to employ the integration in reduced
dimensions with Laguerre polynomials as the interpolant. The
Laguerre polynomials are more suited to the present problem as
they are well defined in the interval $r\in [0,\infty)$ and
satisfy the boundary conditions of the wave function of the GP
equation~\cite{Edwards_Burnett}.

\begin{figure}[!ht]
\begin{center}
\includegraphics[width=0.8\linewidth]{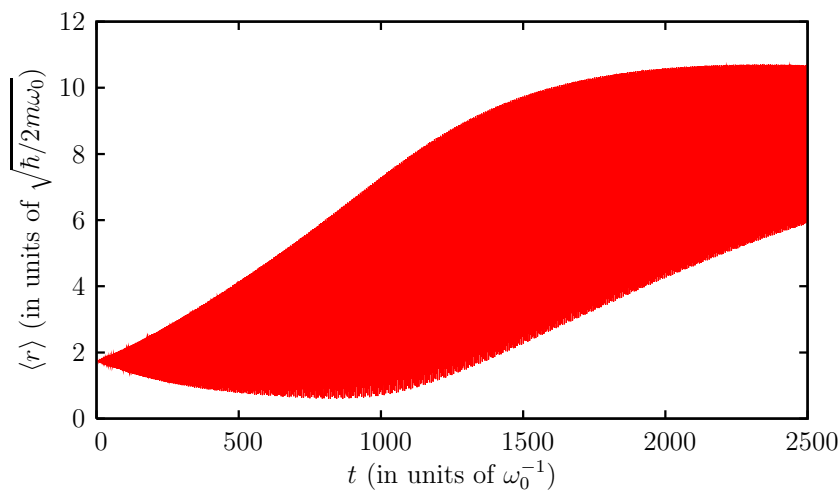}
\caption{ Plot showing the \emph{rms} value $\langle r \rangle$ as
a function of time in the case constant frequency of the temporal
modulation of the scattering length. The parameters are chosen as
$\epsilon_0 = 0$,\, $\epsilon_1 = 0.01$,\, $\mu = 0$ and $\delta =
0$ in Eq.~(\ref{eq4}).} \label{fig:res}
\end{center}
\end{figure}

\begin{figure}[!ht]
\begin{center}
\includegraphics[width=0.8\linewidth]{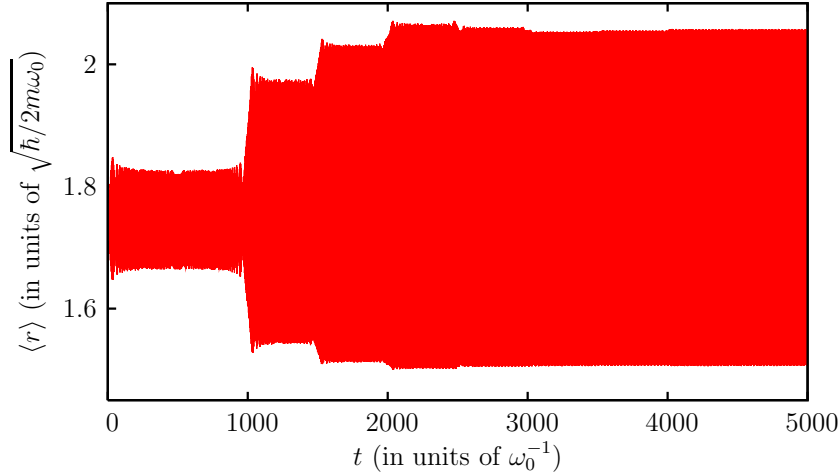}
\caption{ Plot showing the \emph{rms} value $\langle r \rangle$ as
a function of time in the case slowly varying frequency of the
temporal modulation of the scattering length. The parameters are
chosen as $N = 1$,\, $\omega_0 = 1$,\, $\epsilon_0 = 0$,\,
$\epsilon_1 = 0.01$,\, $\mu = 2 \times 10^{-3}$ and $\delta = 0$
in Eq.~(\ref{eq4}).} \label{fig:ares}
\end{center}
\end{figure}

We consider the case of periodically varying scattering length which
can be achieved by means of Feshbach resonance. In order to take
into account the temporal modulation the scattering length can be
taken as
\begin{subequations}
\label{eq4}
\begin {align}
\tilde a(t) &= \epsilon_0 + \epsilon_1 \cos\left[\int \omega
(t)\,dt\right],
\end {align}
and
\begin{align}
 g(t) = \frac{N\tilde a(t)}{l},
\end{align}
where
\begin {align}
\omega(t) &= 2 \omega_0 + \delta - \mu t.
\end {align}
\end{subequations}
is the frequency of the applied field with $\mu \ll 1$ and $\delta$
is a constant. We choose the parameter, without loss of generality
and within the critical threshold limit for collapse, $N=1$. The
trap frequency $\omega_0$ is fixed at $1$. In order to study the
resonance dynamics of (\ref{gpe_r}), we calculate the root mean
square distance, $\langle r \rangle = \sqrt{\langle r^2 \rangle}$,
which is defined as
\begin{align}
\langle r^2 \rangle = \frac{1}{2 \sqrt{2}}\int_0^{\infty} r^2\,
\vert \varphi(r,t) \vert^2\, dr \label{rrms}.
\end{align}
When $\delta = 0$ and $\mu = 0$, the above system (\ref{gpe_r})
exhibits nonlinear resonance in which the steady growth of the
$\langle r \rangle$ is arrested by the nonlinear frequency of the
system~\cite{Ska_pm_jpb:2}. Fig.~\ref{fig:res} shows the plot of
$\langle r \rangle$ as a function of time obtained by solving
(\ref{gpe_r}) numerically for $\delta = 0$ and $\mu =0$.
Fig.~\ref{fig:res} clearly indicates that the steady growth of
$\langle r \rangle$ is controlled by the nonlinearity. However, for
$\delta = 0$ and $\mu = 2\times 10^{-3}$, one observes an
interesting resonance phenomenon giving rise to a step-wise increase
in $\langle r \rangle$ as illustrated in Fig.~\ref{fig:ares}. To
understand such resonance excitations, we study the GP equation
(\ref{gpe_r}) by reducing it to a set of ODEs using a variational
method and then by applying a canonical perturbation theory.

\section{Variational procedure}
\label{variational}

The variational method is one of the simplest way to analyze  the
dynamics of Bose-Einstein
condensate~\cite{Saito_ueda:prl2003,Anderson_pra1983:1,Abdul_etal:phyd2003}.
In this method the original GP equation (\ref{gpe_r}) is reduced to
a system of ODEs, with fewer variables describing the condensate
wave packet, by assuming suitable trial
wavefunction~\cite{Saito_ueda:prl2003,Anderson_pra1983:1,Abdul_etal:phyd2003}.
For convenience, Eq.~(\ref{gpe_r}) can be rewritten as (using the
definition $\varphi(r,t) = r \phi(r,t)$)
\begin{align}
i\frac{\partial \phi}{\partial t} = -\frac{\partial^2 \phi}
{\partial r^2} - \frac{2 }{r} \frac{\partial \phi}{\partial
r}+\frac{r^2}{4} \phi + g(t)\vert\phi\vert^2\phi.
\end{align}
According to the variational method we assume the Gaussian wave
function in the form
~\cite{Saito_ueda:prl2003,Abdul_etal:pra2003:1,Ska_pra:04-01,Anderson_pra1983:1,Abdul_etal:phyd2003,Ska_pre:04-01}
\begin{align}
\phi(r,t)=A(t) \exp\left[-\frac{r^2 }{2a(t)^2}+\frac{ib(t)^2
r^2}{2} +i\delta(t)\right],\label{gaussian}
\end{align}
where $A(t)$, $a(t)$, $b(t)$ and $\delta(t)$ are amplitude, width,
chirp and phase, respectively\cite{Abdul_etal:phyd2003}, which are to
be determined.

The equation for the wave packet parameters $A(t)$, $a(t)$, $b(t)$
and $\delta(t)$ can be obtained by calculating the average
Lagrangian as~\cite{Abdul_etal:phyd2003}
\begin{align}
L(t)=4\pi\int_0^\infty r^2 {\mathcal L}(r,t) dr,
\end{align}
where
\begin{align}
{\mathcal L}(r,t) = \frac{i}{2}(\phi_t^*\phi-\phi_t\phi^*) +
\vert\nabla\phi\vert^2 + \frac{r^2}{4}\vert\phi\vert^2 +
\frac{g(t)}{2}\vert\phi\vert^4
\end{align}
is the Lagrangian density.

One can easily see that the above Gaussian  form assumption
(\ref{gaussian}) leads to the Lagrangian density,
\begin{align}
{\mathcal L}(r,t) = &\, \Bigg[A^2r^2\left(\frac{\dot{b}}{2}
+\frac{1}{a^4}+b^2+\frac{1}{4}\right) 
 +A^2\dot{\delta}\Bigg]e^{-r^2/a^2} + \frac{g(t)}{2}A^4
e^{-2r^2/a^2}
\end{align}
and the average Lagrangian is given by
\begin{align}
L(t)=&\, \frac{3\pi^{3/2}}{2}A^2 a^5 \left(\frac{\dot{b}}{2}
+\frac{1}{a^4}+b^2+\frac{1}{4}\right) 
+\pi^{3/2}A^2 a^3\dot{\delta}
+\frac{\pi^{3/2}}{8\sqrt{2}}g(t)A^4\frac{a^3}{4 }.
\end{align}
The normalization condition for the wave
function $\phi(r,t)$ as
\begin{align}
\int_0^{\infty} r^2 \vert \phi(r,t) \vert^2\,dr = 2 \sqrt{2}.
\end{align}
From the above condition we get $A^2 = \frac{8 \sqrt{2}} {a^3
\sqrt{\pi}}$. Therefore the average Lagrangian density can be rewritten as
\begin{align}
L(t)=&\frac{16 \sqrt{2 \pi} g(t)}{ a^3}
+\frac{12 \sqrt{2}\pi }{a^2}+3 \sqrt{2}\pi a^2 
+ 12 \sqrt{2} \pi b^2a^2+6 \sqrt{2} \pi a^2\dot{b}+ 8 \sqrt{2} \pi \dot{\delta}.
\end{align}
Then the Euler-Lagrangian equations for the Lagrangian $L(t)$ lead
to the following equations for the width $a(t)$ and chirp $b(t)$:
\begin{align}
a_t = 2ab,\;\; b_t=\frac{ 4 g(t)}{\sqrt{\pi} a^5} + \frac{2}{a^4} - 2b^2
- \frac{1}{2}. \label{eqn:at}
\end{align}
On eliminating $b$ from the above
Eq.~(\ref{eqn:at}), one is lead to the following evolution equation
for $a$,
\begin{align}
a_{tt}+a=\frac{4}{a^3} + \frac{8 g(t)}{\sqrt{\pi} a^4}. \label{eqn:ode}
\end{align}
The above equation (\ref{eqn:ode}) can be rewritten as
\begin{align}
a_{tt}+a=\frac{4}{a^3} + \frac{\epsilon(t)}{a^4}, \label{eqn:width}
\end{align}
where
\begin{align}
\epsilon(t) = &\frac{8 N}{\sqrt{\pi}}\left ( \epsilon_0 + \epsilon_1 \cos
\left[\int \omega (t)\, dt\right] \right ),\\
\omega(t) = & 2 \omega_0 + \delta - \mu t.
\end{align}
For $\epsilon(t) = 0$, Eq. (\ref{eqn:width}) is nothing but the well
known Pinney equation \cite{Pinney:1950}. This equation is
completely solvable and the solution takes the following form
\cite{Vkc_proc:2006}
\begin{align}
a(t)=\left[A+\sqrt{A^2-4} \sin(2 t+\delta)\right]^\frac{1}{2}\label{pinney}.
\end{align}
Here $A$ and $\delta$ are arbitrary constants. The above nonlinear
system Eq. (\ref{eqn:width}) with $\epsilon(t) = 0$ possesses the
interesting dynamical property that the frequency of the system is
completely independent of the amplitude, unlike the case of standard
nonlinear oscillators. For exceptions, see ref. \cite{Vkc:2005}.
However, for $\epsilon(t) \neq 0$ one has to study the full system
Eq. (\ref{eqn:width}) to understand the dynamics.
\begin{figure}[!ht]
\centering
\includegraphics[width=0.8\linewidth]{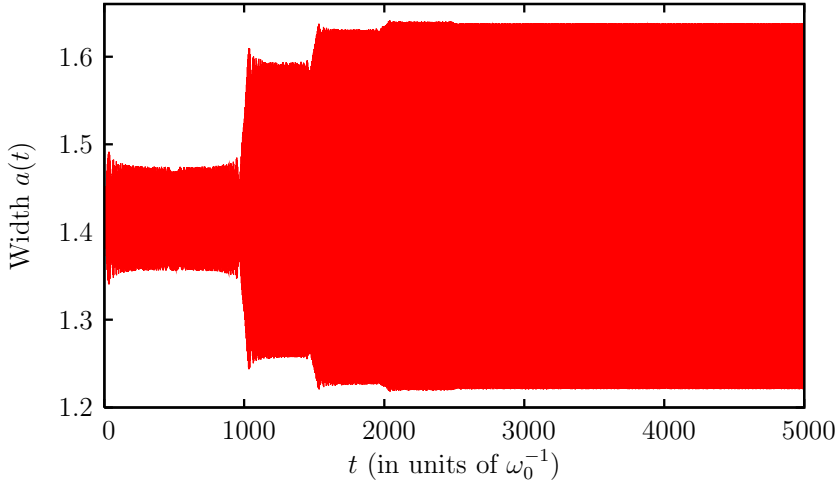}
\caption{ Plot showing the width $a$ as a function of time
calculated from equation (\ref{eqn:width}). The parameters are $N
= 1$,\, $\omega_0 = 1$,\, $\epsilon_0 = 0$,\, $\epsilon_1 =
0.01$,\, $\mu = 2 \times 10^{-3}$ and $\delta = 0$.}
\label{fig:vari}
\end{figure}

In order to study the nature of oscillation for the width $a$ we
solve the above equation (\ref{eqn:width}) numerically using a
fourth order Runge-Kutta method with appropriate initial conditions.
For $\mu \neq 0$\, $(\mu = 0.002)$, we observe that a step-wise
increase in the amplitude of oscillation in $a$ as shown in
Fig.~\ref{fig:vari}. However, when $\mu = 0$ this shows a usual
nonlinear resonance phenomena~\cite{Abdul_etal:phyd2003}.

One can easily relate the width $a(t)$ of the wavefunction
obtaining from the the solution of the ODE (16) to the root mean
square distance of the condensate $\langle r \rangle$ as
\begin{align}
\langle r \rangle = \left[ \frac{1}{2 \sqrt{2}}\int_0^{\infty}
r^4\, \vert \phi(r,t) \vert^2\, dr\right]^{\frac{1}{2}}, \nonumber
\end{align}
which can be evaluated using the from of the wavefunction
$\phi(r,t)$ [eq.~(7)]  and the fact that the amplitude of the
wavefunction is given by the relation $A^2 = \displaystyle \frac{8
\sqrt{2}} {a^3\sqrt{\pi}}$ obtained from the normalization
condition (12). Thus the width of the condensate can be related to
the root mean square distance of the condensate as $\langle r
\rangle = a \sqrt{\frac{3}{2}} $.

\section{Perturbation theory in the action-angle variables}
\label{action}

In order to have a more clear understanding on the resonance
phenomena, we analyze the Eq.~(\ref{eqn:width}) by constructing the
action-angle variables.Treating Eq.~(\ref{eqn:width}) as nearly
integrable system($\epsilon(t) \ll 1$), it is more convenient to
consider the action-angle variables in which one can employ a
perturbation theory \cite{Lichtenberg_lieberman:1983}.

\subsection{The action-angle variables for the unperturbed system}

First, let us consider the unperturbed system $\epsilon(t) =0$. The
Hamiltonian for the unperturbed problem is given by
\begin{align}
H_0=\frac{a_t^2}{2}+U(a),
\end{align}
where
\begin{align}
U(a)=\frac{a^2}{2}+\frac{2}{a^2} \notag
\end{align}
If $E$ is the energy then the width, $a(t)$, oscillates between
$a_{\mbox{\scriptsize min}}=\sqrt{E-\sqrt{E^2-4}}$ and
$a_{\mbox{\scriptsize max}}=\sqrt{E+\sqrt{E^2-4}}$. Since
the Hamiltonian is conserved, i.e., $H_0 = E$,
\begin{align}
\int \frac{da}{\sqrt{2E-a^2-4a^{-2}}} = \int dt.
\end{align}
On integrating the above equation, we obtain
\begin{align}
a(t)=\sqrt{E+\sqrt{E^2-4}\, \sin \theta}  \label{eqn:ode4},
\end{align}
where $\theta = \theta_0 + 2t$ and $\theta_0$ being integration
constant, see Eq. (\ref{pinney}).

We now make  a change to action-angle variables. In this problem,
where the phase space is two-dimensional, there is one action
variable $I$ and a conjugate angle variable $\theta$. The action
variable is given by
\begin{align}
I=&\, \frac{1}{2\pi}\oint a_t da\notag \\
 =& \, \frac{1}{\pi}
\int_{a_{\mbox{\scriptsize min}}}^{a_{\mbox{\scriptsize max}}}
{\sqrt{2E-a^2-\frac{4}{a^{2}}}} \; da.
\end{align}
This integral can be easily evaluated by using the energy ($E$) to express
the momentum $a_t$ in terms of position $a$. The resulting integral is
elementary and leads to
\begin{align}
I=\frac{1}{2}(E-2).
\end{align}
By expressing the total energy $E$ in terms of the action $I$,
the unperturbed Hamiltonian is written as
\begin{align}
H_0 = E =2(1+I),\label{uhxp}
\end{align}
which is a function of $I$ only. The change of variables to
action-angle variable is canonical, so that the Hamiltonian's
equations retain their form
\begin{subequations}
\begin{align}
\frac{dI}{dt} & = -\frac{\partial H_0(I)}{\partial
\theta}=0, \\
\frac{d\theta}{dt} & =\frac{\partial H_0(I)}{\partial
I}=\Omega_0,\label{uhem}
\end{align}
\end{subequations}
where $\Omega_0 = 2$.

\subsection{The action-angle variables for the system with
perturbation}

Now we consider the problem with perturbation. Then the Hamiltonian for
perturbed problem (\ref{eqn:width}) can be written as
\begin{align}
H = H_0+\epsilon(t) H_1 \equiv H_0+\epsilon(t) \frac{1}{ 3 a^3}, \label{hxp}
\end{align}
where
\begin{align}
\epsilon(t)= \left(\frac{8 N \epsilon_1}{\sqrt{\pi}}\right)\cos
(2t+\delta t - \mu t^2),\;\;\;\;\
\mbox{and}\;\;\epsilon_0 \notag
= 0.
\end{align}
Further, one can express $H$ in terms of action angle variables
$I$ and $\theta$ using the Eqs.~(\ref{eqn:ode4}) and
(\ref{uhxp}),
\begin{align}
H(I,\theta) = H_0(I)+\epsilon(t) H_1(I,\theta), \label{ham:perturb}
\end{align}
where
\begin{align}
H_1 (I,\theta)= \, & \frac{1
}{ 6 \sqrt{2} \left(1+I+\sqrt{I(2+I)}
 \sin\theta \right)^{3/2}}, \label{hxp2}
\end{align}
\begin{figure}[!ht]
\centering
\includegraphics[width=0.8\linewidth]{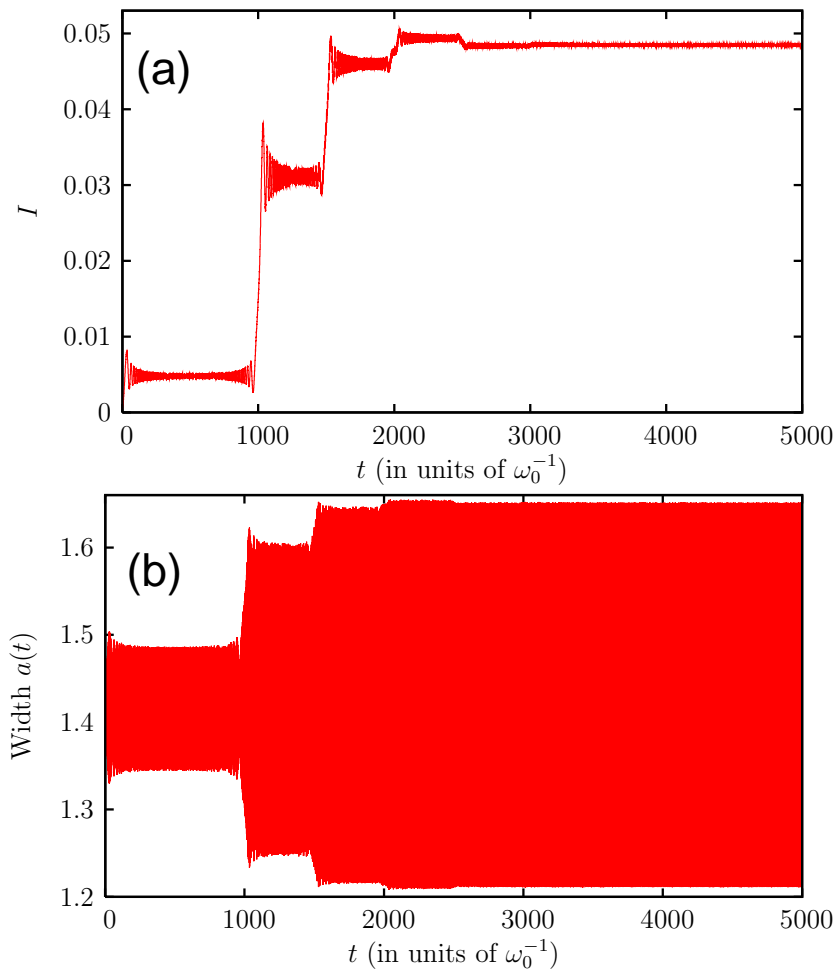}
\caption{Plot showing (a) the action $I$ as a function of time
obtained by solving the reduced system Eq.~(\ref{hxp8}) and (b)
the width $a(t)$ obtained from eq.~\ref{width} as a function of
$t$. The parameters are $N = 1$,\, $\omega_0 = 1$,\, $\epsilon_0 =
0$,\, $\epsilon_1 = 0.01$,\, $\mu = 2 \times 10^{-3}$ and $\delta
= 0$.} \label{fig:action}
\end{figure}
Thus the Hamiltonian equations of motion are given by
\begin{subequations}
\begin{align}
\frac{dI}{dt} &  = -\epsilon(t) \frac{\partial H_1}{\partial \theta}, \\
\frac{d\theta}{dt} & = \Omega_0 + \epsilon(t) \frac{\partial H_1}{\partial I}
\end{align}
\label{hxp8}
\end{subequations}
where
\begin{align}
\frac{\partial H_1}{\partial \theta}= -\frac{\sqrt{I(2+I)}\,
\cos\theta}{4\sqrt{2}\left(1+I+\sqrt{I(2+I)}\sin\theta\right)^{5/2}},
\notag
\end{align}
\begin{align}
\frac{\partial H_1}{\partial I}=
-\frac{1+\displaystyle\frac{1+I}{\sqrt{I(2+I)}}\, \sin
\theta }{ 4 \sqrt{2}\left(1+I+\sqrt{I(2+I)}\sin\theta\right)^{5/2}}. \notag
\end{align}
The values of $I$ and $\theta$ are obtained, as a function of $t$,
by solving Eqs.~(\ref{hxp8}) numerically using a fourth order
Runge-Kutta method with appropriate initial conditions.
Fig.~\ref{fig:action}(a) shows the plot of $I$ as a function of
time. It is evident that $I$ retains relatively small values.
Substituting $I$ and $\theta$ values in equations (\ref{uhxp}) and
(\ref{eqn:ode4}), we get the expression for condensate width $a$
in terms of action-angle variables as
\begin{align}
a=\left[ 2(1+I)+2 \sqrt{I(2+I)}\sin\theta\right]^\frac{1}{2}.\label{width}
\end{align}
We note that the variation of $a$ [in Eq.(\ref{width})] depicted
in Fig.~\ref{action}(b), after using the results for $I$, compares
well with that shown in Fig.~\ref{fig:vari}.
\begin{figure}[!ht]
\begin{center}
\includegraphics[width=0.8\linewidth]{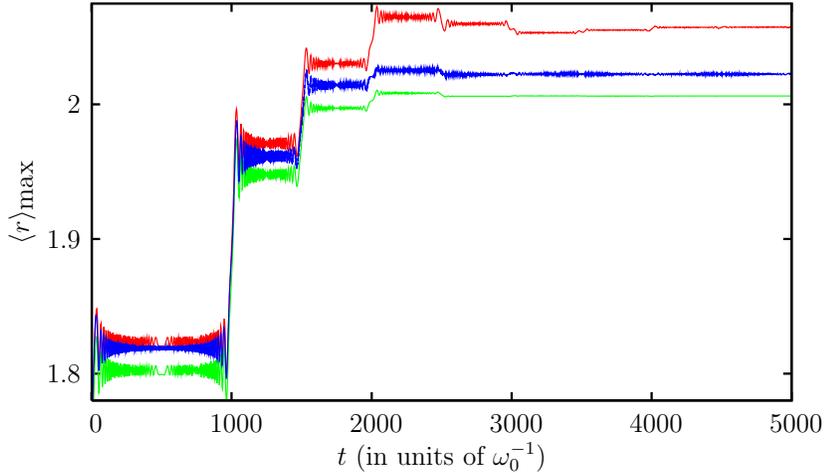}
\caption{Plot of the $\langle r \rangle_{max}$ obtained by solving
eq.~(\ref{gpe_r}), from the numerical solutions of
eq.~(\ref{eqn:ode}) and from perturbation theory (\ref{hxp8}.}
\label{fig:comp}
\end{center}
\end{figure}
Further we have also compared, in Fig.~\ref{fig:comp}, the results
obtained from the direct numerical solution of the GP equation
(\ref{gpe_r}), using the numerical solution of (\ref{eqn:ode})
obtained from variational procedure and the results of the
perturbation theory by solving equations (\ref{hxp8}).

\section{Canonical perturbation Theory}
\label{canonical}

In order to understand the resonance phenomena one has to find out
the approximate nonlinear frequency of the above system which is
responsible for the resonant excitations. For this purpose we adopt
a canonical perturbation theory to find the nonlinear frequency of
near integrable
systems~\cite{Lichtenberg_lieberman:1983,Tabor:1989,Lakshmanan_rajasekar:2003}.
The basic idea of this method is to rewrite the perturbed system
(\ref{ham:perturb}) into a new set of action angle variables
$(J,\phi)$ by a canonical transformation to a new Hamiltonian
$K(J)$, which depends on the new action $J$ only, that is,
\[
H(I,\theta) \rightarrow K(J).
\]
For  $I \ll 1$, one can express the perturbed Hamiltonian $H_1$
in the binomial expansion form as,
\begin{align}
H_1(I,\theta) \approx \frac{1}{6 \sqrt{2}}
 \left[ 1 -\frac{3}{2}\left(I + \sqrt{I(2+I)}\sin\theta \right)\right]
\end{align}
We now attempt a canonical transformation to action-angle variables
($J$,$\phi$) for $H(I,\theta)$ through a type II generating
function~\cite{Goldstein:1980} as
\begin{align}
S(J,\theta)=S_0(J,\theta)+\epsilon(t) S_1(J,\theta)+\epsilon(t)^2
S_2(J,\theta)+\ldots,\label{hxp6}
\end{align}
where $S_0 =J \theta$ is the identity generator. Then the new
action-angle variables are given by~\cite{Goldstein:1980}
\begin{subequations}
\label{hxp3}
\begin{align}
I &=\frac{\partial S(J,\theta)}{\partial \theta},  \\
 \phi & = \frac{\partial S(J,\theta)}{\partial J}.
\end{align}
\end{subequations}
Rewriting the Hamiltonian (\ref{ham:perturb}) in terms of $J$ using the
relation~(\ref{hxp3}), such a transformation gives the new
Hamiltonian $K(J)$, that is,
\begin{align}
H_0\left(\frac{\partial S(J,\theta)}{\partial \theta}\right)+
\epsilon(t) H_1\left(\frac{\partial S(J,\theta)}{\partial
\theta},\theta\right) = K(J).\label{hxp7}
\end{align}
Furthermore, one now expands the new Hamiltonian $K(J)$ as a
power series of $\epsilon(t)$, that is,
\begin{align}
K(J)=K_0(J)+\epsilon(t) K_1(J)+\epsilon(t)^2 K_2(J)+\ldots\label{hxp5}
\end{align}
By making a Taylor's series expansion of $H_0$ and $H_1$ in
Eq.~(\ref{hxp7}), using the relation for $S(J,\theta)$ from (\ref{hxp6}),
and equating the coefficients of various powers of $\epsilon(t)$, one
can obtain a system of equations. The new Hamiltonian $K(J)$ can then be
obtained by solving this set of equations.

The zeroth order term $K_0(J)=H_0$ of the new Hamiltonian is
obtained by  equating the coefficient of $\epsilon(t)^0$. It is easy
to see that $K_0(J)$ can be simply found by replacing $I$ by $J$ in
the zeroth order Hamiltonian $H_0$. Therefore the zero-order
frequency is given by,
\begin{align}
\Omega_0 =\frac{\partial K_0}{\partial J} =2 \label{fre0}
\end{align}
Similarly, equating the coefficient of $\epsilon(t)$, one obtains
\begin{align}
K_1(J) =\, & \frac{\partial S_1}{\partial \theta}\frac{\partial
H_0}{\partial J} H_1(J,\theta) \notag \\ = \, & \Omega_0
\frac{\partial S_1}{\partial \theta} H_1(J,\theta)\label{hxpk1}
\end{align}
At this point, one can exploit the periodicity of the motion in the
angle variable $\theta$. Since $S_i$, $i=1, 2, \ldots$, are assumed
to be periodic in $\theta$. The averaging of Eq.~(\ref{hxpk1}) over
$\theta$ leads to the mean of the derivatives of $S_i$ vanish. Then
the first-order Hamiltonian correction is given by
\begin{align}
K_1(J) =\, &1-\frac{3}{2}J.
\end{align}
Hence the first order correction to frequency is given by
\begin{align}
\Omega_1(J) & = \frac{\partial K_1}{\partial J} \notag\\
& = -\frac{3}{2}. \label{fre1}
\end{align}
from Eq.~(\ref{hxp3}) upto first order,
\begin{subequations}
\label{hxp9}
\begin{align}
J &=I-\epsilon(t) \frac{\partial S_1(I,\theta)}{\partial \theta}, \\
\phi &=\theta+\epsilon(t) \frac{\partial
S_1(I,\theta)}{\partial \theta}.
\end{align}
\end{subequations}
We get $I$ and $\theta$ values by solving Eq.~(\ref{hxp8})
numerically and substituting these values in Eq.~(\ref{hxp9}), we
get the $J$ value.
\begin{figure}[!ht]
\begin{center}
\includegraphics[width=0.8\linewidth]{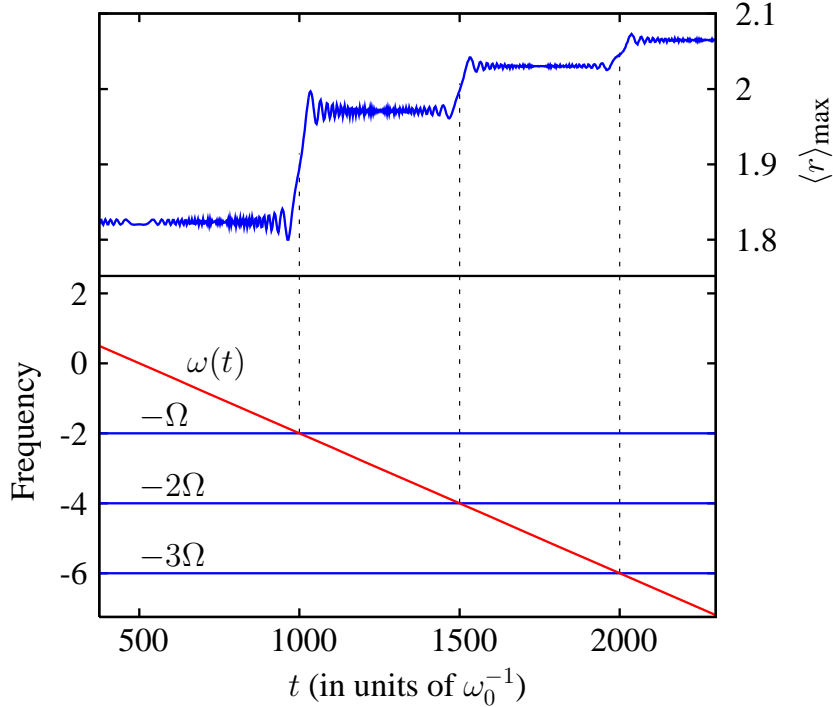}
\caption{ Phase locking between nonlinear system frequency and
driven frequency. (top) Plot of the $\langle r \rangle_{max}
$obtained by solving Eq.~(\ref{gpe_r}) and (bottom) the driving
frequency and the different harmonics of nonlinear frequency is
shown.} \label{fig:freq}
\end{center}
\end{figure}

The total nonlinear system frequency is,
\begin{align}
\Omega(J) =\Omega_0(J)+\epsilon(t) \Omega_1(J)
\notag.
\end{align}
In the above expression the second term  in the right hand side can
be neglected as $\epsilon(t) $ is very small and one can find
$\Omega(J) \approx\Omega_0(J)$. Here the nonlinear frequency is
almost constant while the driving  frequency is slowly varying with
time. Hence one could expect a phase locking when the driving
frequency matches with the nonlinear frequency or its integer
multiples.

From Fig.~\ref{fig:freq} it is easy to see that at $t \approx
1000$, the driving frequency $\omega(t)$ matches with the
nonlinear system frequency $\Omega$ (i.e, $\omega(t)=-\Omega$), at
which the first resonance jump occurs in the original problem as
shown in Fig.~\ref{fig:ares}. It explains the primary resonance
that occurs at $t \approx 1000$ in the original problem
Eq.(\ref{gpe_r}). At another time, $t \approx 1500$, where
$\omega(t) = -2 \Omega$, a subharmonic resonance occurs leading to
a second jump in the amplitude as shown in Fig.~\ref{fig:ares}. As
a result a step-wise increase in the amplitude  as shown in
Fig.~\ref{fig:ares} due to the phase locking between drive and
system whenever the driving frequency matches with the system
nonlinear frequency or its integral multiples (ie, $\omega(t)= n
\Omega$, where $n=0,\pm 1, \pm 2,\ldots$). We refer this phenomena
as a kind of subharmonic autoresonance.

\section{Summary and conclusions}
\label{conclusion}

In summary, we have studied the autoresonant excitations in
Bose-Einstein condensates under the action of external periodic
modulation with time dependent frequency. By numerically solving the
corresponding Gross-Pitaevskii equation we have observed that there
occurs a successive phase locking with step-wise increase in the
overall amplitude of oscillation. We have employed a variational
procedure using Gaussian trial wave function to simplify the problem
in fewer coordinates. Then the reduced system has been found to show
a kind of subharmonic autoresonance phenomenon with successive
resonance excitations. Further, we have studied the problem in the
action-angle variables which allows one to make a perturbation
analysis and thereby identify the approximate frequency that is
responsible for the successive resonant (step-wise) excitations. The
results obtained from the canonical perturbation theory compare well
with the numerical solution of the original problem.

\section*{Acknowledgments}

This work is supported in part by the Department of Science and
Technology (DST) and National Board for Higher Mathematics
(Department of Atomic Energy) Government of India.

\end{document}